\documentclass{article}
\usepackage{ijcai26}

\usepackage{times}
\usepackage{soul}
\usepackage{url}
\usepackage[hidelinks]{hyperref}
\usepackage[utf8]{inputenc}
\usepackage[small]{caption}
\usepackage{graphicx}
\usepackage{amsmath}
\usepackage{amsthm}
\usepackage{amsfonts}
\usepackage{booktabs}
\usepackage{algorithm}
\usepackage[noend]{algorithmic}
\floatname{algorithm}{Protocol}

\usepackage[switch]{lineno}
\usepackage{comment}
\excludecomment{dontshow}
\usepackage{colortbl}
\usepackage{xcolor}

\newtheorem{example}{Example}
\newtheorem{theorem}{Theorem}

\usepackage{stmaryrd}

\newcommand{\bigO}{\mathcal{O}}

\newcommand{\mB}{\mathcal{B}}
\newcommand{\mC}{\mathcal{C}}

\newcommand{\mF}{\mathcal{F}}
\newcommand{\mP}{\mathcal{P}}

\newcommand{\mV}{\mathcal{V}}

\newcommand{\score}{\textrm{score}}
\newcommand{\tsigma}{\tilde{\sigma}}
\newcommand{\trho}{\tilde{\rho}}
\newcommand{\tR}{\tilde{R}}

\usepackage{ifthen}
\newboolean{isIJCAI}
\setboolean{isIJCAI}{false} 
\newcommand{\showIJCAIorHAL}[2]{%
  \ifthenelse{\boolean{isIJCAI}}{#1}{#2}%
}

\title{The Communication Complexity of Instant-Runoff Voting}

\author{
\'Elie de Panafieu$^{1}$\and
Fran\c{c}ois Durand$^1$ \and
J\'er\^ome Lang$^2$ \\
\affiliations
$^1$Nokia Bell Labs\\
$^2$CNRS, LAMSADE, Universit\'e Paris-Dauphine, PSL, France\\
\emails
\{francois.durand, elie.de\_panafieu\}@nokia-bell-labs.com,
lang@lamsade.dauphine.fr
}

\begin{document}

\maketitle

\begin{abstract}
The communication complexity of a voting rule is the worst-case number of bits that $n$ voters must transmit to a central authority under the most efficient elicitation protocol in an election with $m$ candidates.
We study the communication complexity of Instant-Runoff Voting (IRV).
Conitzer and Sandholm [2005] established an upper bound of $O(n (\log m)^2)$, but did not provide a matching lower bound beyond $\Omega(n \log m)$.
We resolve this open problem by raising the lower bound to $\Omega(n (\log m)^2)$ using the fooling set technique, thereby showing that the communication complexity of IRV is $\Theta(n (\log m)^2)$.
We further show that this complexity drops to $\Theta(n \log m)$ under the single-peakedness restriction, and that both the IRV-Average variant and Single Transferable Vote (STV), the multiwinner extension of IRV, have the same asymptotic communication complexity as~IRV.
\end{abstract}

\showIJCAIorHAL{%
An \href{https://shs.hal.science/hal-05566718}{\textcolor{blue}{extended version with appendices}} and a \href{https://youtu.be/gTXV3R2DS6o}{\textcolor{blue}{short video presentation}} are available online \cite{hal_version,video}.%
}{%
A \href{https://youtu.be/gTXV3R2DS6o}{\textcolor{blue}{video}} presentation is available online~\cite{video}.%
}

\section{Introduction}

Assessing the suitability of a voting rule requires analyzing its normative properties, its computational issues, and its \emph{communication complexity}, while also taking into account the rule’s understandability by laypeople.
Among these criteria, communication complexity is especially relevant: if excessive interaction is required, voters may abandon the voting process altogether or, at best, submit their preferences hastily and without due care.
In this respect, communication complexity is arguably more directly relevant to practical deployment than computational complexity: the cost of computation falls on a machine, whereas the cost of communication falls on voters.

A communication protocol for a voting rule is an interactive elicitation procedure in which, at each step, some voters are asked to transmit information about their vote, conditional on the responses obtained in previous steps, and that fully determines the outcome of the election.
The complexity of a communication protocol is the total number of bits transmitted by voters in the worst-case execution of the protocol.
The communication complexity of a voting rule is the minimum complexity over all its communication protocols.

In this paper, our main focus is on the communication complexity of \emph{Instant-Runoff Voting} (IRV), a prominent single-winner voting rule used in many large-scale political elections.
IRV proceeds by successive elimination of the candidate with the fewest top-ranked votes.
The rule is sometimes also referred to as \emph{Single Transferable Vote} (STV); following standard terminology, we use IRV for the single-winner rule and STV for its multiwinner variant.

The communication complexity of voting rules was first, and mostly, studied by Conitzer and Sandholm~\shortcite{ConitzerS05}. In this framework, only the bits sent by voters are counted: messages sent by the central authority and received by voters are not taken into account, as the communication (or elicitation) burden falls on voters rather than on the central authority, which is typically a computer.
Their upper bounds are established via explicit communication protocols, whereas their lower bounds rely on the classical \emph{fooling set} technique \cite{KushilevitzN97}, which we recall later.
Their results identify the communication complexity of most common voting rules, with a notable exception: IRV, for which they leave a gap between a lower bound $\Omega(n \log m)$ and an upper bound $O(n (\log m)^2)$. To date, this gap has not been filled yet.

Answering this open question is particularly relevant given the prevalence of IRV in real-world elections, both in large-scale political contexts and in lower-stakes settings. If the communication complexity of IRV is $\Theta(n (\log m)^2)$, then the known upper bound is asymptotically optimal, and efforts should focus on refining and implementing the existing protocol on interactive voting platforms. If, on the other hand, the communication complexity of IRV has a smaller order of magnitude than $n (\log m)^2$, this would imply the existence of a more efficient protocol yet to be discovered.

We do not resist the temptation to spoil the reader: we
close this gap by proving that the communication complexity of IRV is indeed $\Theta(n (\log m)^2)$.

Section~\ref{sec:background+related} provides background and related work.
Section~\ref{sec:framework} introduces the framework.
Section~\ref{sec:ccirv} presents our main result: the communication complexity of IRV is bounded below by $\Omega(n (\log m)^2)$.
In Section~\ref{sec:other} we show that the communication complexity of IRV drops to $\Theta(n \log m)$ under the restriction to single-peaked profiles, and that both the IRV-Average variant and the multiwinner rule STV have the same asymptotic communication complexity as IRV.
Section~\ref{sec:discussion} concludes.

\section{Background and Related Work}\label{sec:background+related}

\subsection{Instant-Runoff Voting (IRV)}\label{background-irv}

IRV is used in a wide range of elections across several countries, including Australia, India, Ireland, the UK, and the US.
It is one of the few commonly studied single-winner rules that are independent of clones \cite{Tideman87}, and arguably the only one used in large-scale real-world elections.
It is computationally hard to manipulate, even by a single voter \cite{BartholdiToveyTrick1989}; the probability that a preference profile is manipulable is low in most settings \cite{durand2023coalitional,Durand25}; and manipulation is cognitively demanding, which tends to encourage sincere voting \cite{StraetenLSB10}.
From a computational perspective, while IRV with immediate tie-breaking is easy to compute, its \emph{parallel universe} version is NP-hard, though still tractable in practice \cite{CsarLPS17,WangSSZJX19}.
IRV has also been the subject of numerous empirical studies: see \cite{empiricirv} for a continuously updated survey.

\subsection{Communication Complexity of Voting Rules}\label{background-cc}

The seminal paper by Conitzer and Sandholm~\shortcite{ConitzerS05} identifies three broad classes of voting rules.
\emph{High-communication} rules, such as Borda, Copeland, and ranked pairs, require voters to transmit their full rankings in the worst case, yielding a communication complexity of $\Theta(n m \log m)$.
At the other end of the spectrum, \emph{low-communication} rules, such as plurality and plurality with runoff, require each voter to transmit not much more than a single candidate, leading to a communication complexity of $\Theta(n \log m)$.
\emph{Medium-communication} rules lie between these extremes; notable examples include voting trees and Bucklin, both of which have communication complexity $\Theta(n m)$.
For IRV, the gap between $\Omega(n \log m)$ and $O(n (\log m)^2)$ prevents classification as a medium-communication rule, although it is widely believed to require substantially more communication than $n \log m$.

Service and Adams~\shortcite{ServiceA12} study the communication complexity of approximating voting rules.
Mandal \emph{et al.}~\shortcite{MandalP0W19} investigate the trade-off between communication complexity and distortion.
Chevaleyre \emph{et al.}~\shortcite{ChevaleyreLMR09}, Xia and Conitzer~\shortcite{XiaC10}, and Karia and Lang~\shortcite{karia2024compiling} focus on the one-round communication complexity of voting rules, known as \emph{compilation complexity}.

\subsection{Communication Issues for IRV}\label{cc-irv}

The raw definition of IRV assumes full rankings and therefore entails a high communication burden. Two natural escape routes are \emph{simplified ballots} and \emph{interactive protocols}.

The most common form of simplified ballots consists of \emph{truncated ballots}: voters rank only their top $k$ candidates, where $k$ may depend on the voter.
Smaller values of~$k$ make the input easier to elicit but provide weaker guarantees that the resulting winner coincides with the true IRV winner.
Such truncated-ballot approximations of IRV and STV have been studied both theoretically and empirically \cite{BurnettKogan15,AyadiBLP19,KilgourGF20,HoffmannKRW21,TomlinsonUK23}.
Other works allow voters to express indifference \cite{DelemazureP24}, or consider missing and truncated ballots \cite{HanMX24}.
All these works assume one-shot elicitation.

By contrast, to guarantee determination of the true IRV winner, one can rely on interactive elicitation protocols.
Such protocols ask voters to provide partial preference information to the central authority as needed during execution.
As mentioned above, Conitzer and Sandholm~\shortcite{ConitzerS05} propose an IRV protocol with communication complexity $\bigO(n (\log m)^2)$.
Ayadi \emph{et al.}~\shortcite{AyadiBLP19} improve its practical performance while preserving the same worst-case complexity.

\section{Framework}
\label{sec:framework}

\subsection{Rankings and Profiles}

We consider a set of $n$ voters $\mV = \{0, \ldots, n-1\}$ and a set of $m$ candidates $\mC = \{0, \ldots, m-1\}$.
A \emph{ranking} $r$ is a linear order over $\mC$, for instance $(2 \succ 0 \succ 1)$.
A \emph{profile} $P$ is a collection of $n$ rankings, one for each voter.
Given two profiles $P$ and $Q$, and a voter $v \in \mV$, we denote by $P^{-v} + Q^{v}$ the profile obtained from $P$ by replacing the ranking of voter~$v$ in $P$ with their ranking in $Q$.

\subsection{Communication Complexity of Voting Rules}

Following \cite{ConitzerS05}, a \emph{voting rule} $f$ maps a profile to a winning candidate.
The \emph{communication complexity} of $f$ is the worst-case number of bits transmitted by the voters in the most efficient protocol implementing $f$.
A \emph{fooling set} $\mF$ is a collection of profiles such that:
\begin{enumerate}
\item $f$ has the same outcome $w$ for all profiles $P \in \mF$; and
\item for any two distinct profiles $P, Q \in \mF$, there exists a profile $T$ obtained by selecting, for each voter $v$, either $P_v$ or $Q_v$, such that $f(T) \neq w$.
\end{enumerate}
Its existence implies a communication complexity of at least $\log_2 |\mF|$ for $f$ \cite{KushilevitzN97}.
As observed by Conitzer and Sandholm~\shortcite{ConitzerS05}, it suffices to construct a fooling set for a representative set of pairs $(n,m)$, namely for infinitely many values of $m$ and, for each such $m$, infinitely many values of $n$.
All these notions extend naturally to multiwinner voting rules, where $f$ outputs a subset of candidates rather than a single winner.

\subsection{Instant-Runoff Voting (IRV)}

Under \emph{Instant-Runoff Voting} (IRV), the outcome is determined through successive rounds of elimination.
Initially, all candidates in $\mC$ are \emph{active}.
In each round, each voter assigns one point to their most-preferred active candidate, and the candidate with the lowest score is eliminated from the active set.
The procedure iterates until a single active candidate remains and is declared the winner.
In practical implementations, it may terminate as soon as a candidate obtains a strict majority of the votes, that is, a score greater than $n/2$, since this candidate is then guaranteed to win.

Throughout the paper, we assume that ties are broken in every round in favor of candidates with smaller indices: if several candidates are tied for the lowest score, the one with the \emph{largest} index is eliminated.
Alternative tie-breaking rules are discussed in Section~\ref{sec:tie_break_trick_irv}.

\section{Communication Complexity of IRV}
\label{sec:ccirv}

Section~\ref{sec:ppr_protocol} recalls the communication protocol for IRV described by Conitzer and Sandholm~\shortcite{ConitzerS05}, yielding an upper bound of $\bigO(n (\log m)^2)$.
To obtain a matching lower bound, Sections~\ref{sec:definition_fooling_set_irv}, \ref{sec:proof_fooling_set_property_irv}, and \ref{sec:cardinality_fooling_set_irv} introduce a family of profiles $\mF$, show that it forms a fooling set, and derive its asymptotic log-cardinality $\Theta(n (\log m)^2)$.
Section~\ref{sec:tie_break_trick_irv}
shows that this bound holds independently of the tie-breaking rule.

\subsection{Progressive Preference Revelation Protocol}\label{sec:ppr_protocol}

Conitzer and Sandholm~\shortcite{ConitzerS05} establish the $\bigO(n (\log m)^2)$ upper bound on the communication complexity of IRV using Protocol~\ref{algo:ppr_irv}, which we call the \emph{Progressive Preference Revelation} (PPR) protocol.
To make it more suitable for real-world usage, we add early termination as soon as a candidate obtains a strict majority of the votes.
Beyond its interest for IRV, this protocol also serves as a basis for analyzing the alternative settings considered in Section~\ref{sec:other}.

\begin{algorithm}[t]
\caption{PPR Protocol for IRV}
\label{algo:ppr_irv}
\begin{algorithmic}
    \STATE $A \leftarrow \mC$ \COMMENT{active candidates}
    \FOR{$v \in \mV$}
        \STATE $t(v) \leftarrow$ \textbf{elicit} the top candidate of $v$
    \ENDFOR
    \WHILE{\TRUE}
        \FOR{$c \in A$}
            \STATE $V(c) \leftarrow \{v \in \mV \mid t(v) = c\}$ \COMMENT{voters supporting $c$}
            \STATE $\score(c) \leftarrow |V(c)|$
        \ENDFOR
        \IF{$\exists\,c \in A \text{ such that } \score(c) > n/2$}
            \RETURN $c$ \COMMENT{$c$ is the winner}
        \ELSE
            \STATE let $c$ be a candidate in $A$ minimizing $\score(c)$ (possibly using tie-breaking)
            \STATE remove $c$ from $A$
            \FOR{$v \in V(c)$}
                \STATE $t(v) \leftarrow$ \textbf{elicit} the top candidate of $v$ in $A$
            \ENDFOR
        \ENDIF
    \ENDWHILE
\end{algorithmic}
\end{algorithm}

\subsection{Definition of the Fooling Set}\label{sec:definition_fooling_set_irv}

To build intuition for our fooling set, we start from perfectly symmetric profiles, where each ranking is represented by the same number of voters.
In such profiles, the elimination order is $(m-1,m-2,\ldots,0)$, since ties are broken in favor of candidates with smaller indices, and the winner is candidate~$0$.

However, some rankings cannot be distinguished by the IRV counting process in such profiles.
For instance, the rankings $(2 \succ 4 \succ 3 \succ 0 \succ 1)$ and $(2 \succ 0 \succ 1 \succ 3 \succ 4)$ both initially count for candidate~$2$ and are then transferred to candidate~$0$, where they remain until the end.
We say that these two rankings have the same \emph{signature} $(2, 0)$.
If two profiles differ only by replacing rankings with others having the same signature, then mixing these profiles cannot alter the IRV outcome and thus cannot prevent candidate~$0$ from winning, contradicting the definition of a fooling set.
Thus, all rankings with the same signature will be replaced by a single ranking, called its \emph{representative}.

Candidates not seen by the IRV counting process, i.e., not appearing in the signature, are ranked so as to favor candidates with higher indices as much as possible.
When mixing two distinct profiles from the fooling set, this choice will help a candidate other than~$0$ to win.

\paragraph{Signature.}

As mentioned above, for a ranking $r$, its \emph{signature} $\sigma(r)$ is the list of candidates observed during the IRV counting process, assuming the elimination order $(m-1, m-2, \ldots, 0)$.
For example, $\sigma\big((2 \succ 4 \succ 3 \succ 0 \succ 1)\big) = (2, 0)$.
By convention, we include a final trivial round in which only candidate~$0$ remains and all voters vote for that candidate.
For example, $\sigma\big((1 \succ 2 \succ 0)\big) = (1, 0)$, even if this ballot supports candidate~1 throughout the entire process in practice.

Formally, the signature $\sigma(r)$ is defined by the following algorithm.
Start with the empty list.
For each rank $i \in \{0, \ldots, m-1\}$, append the candidate $r_i$ if and only if it has the smallest index encountered so far, that is, if $r_i = \min_{j \leq i} r_j$.
In combinatorics, this construction is known as the \emph{record} of a permutation, except that records are conventionally defined with respect to a maximum rather than a minimum \cite{dumont1988developpement}.
The signature of a ranking is therefore always a decreasing list, starting with its top-ranked candidate and ending with candidate~$0$.
The possible signatures are exactly the decreasing lists of candidates ending with~$0$.

\paragraph{Representative.}

Intuitively, for a signature $s$, we define its \emph{representative} $\rho(s)$ as the ranking with signature~$s$ that favors candidates with higher indices as much as possible.
For example, $\rho\big((2, 0)\big) = (2 \succ 4 \succ 3 \succ 0 \succ 1)$.
Formally, $\rho(s)$ is obtained by scanning the signature $s$: for each element $s_i$, we append $s_i$, followed by all candidates with higher indices that have not yet been added, in decreasing order.
By construction, for every signature~$s$, we have $\sigma(\rho(s))=s$.

\paragraph{Fooling set.}

Assume $n = \ell m!$, where $\ell$ is a positive integer.
Let $\mP^{(\ell)}$ denote the set of all profiles in which each ranking appears exactly $\ell$ times.
We define our fooling set $\mF$ as
\[
\mF = \bigl\{\, \rho(\sigma(P_\textrm{ini})) \mid P_\textrm{ini} \in \mP^{(\ell)} \,\bigr\},
\]
where $\rho(\sigma(P_\textrm{ini}))$ denotes the profile obtained from $P_\textrm{ini}$ by replacing each ranking $r$ with $\rho(\sigma(r))$, the representative of its signature.

An equivalent definition is as follows.
Let $R(s) = |\sigma^{-1}(s)|$ denote the number of rankings with signature~$s$.
The fooling set $\mF$ then consists of all profiles that contain, for each signature $s$, exactly $\ell R(s)$ copies of~$\rho(s)$.

For example, for $m = 3$, consider this profile $P_{\textrm{ini}} \in \mP^{(1)}$:
\[
\begin{array}{|c|c|c|c|c|c|}
\hline
\mathbf{0} & \mathbf{0} & \mathbf{1} & \mathbf{1} & \mathbf{2} & \mathbf{2} \\
1        & 2        & \mathbf{0} & 2        & \mathbf{0} & \mathbf{1} \\
2        & 1        & 2        & \mathbf{0} & 1        & \mathbf{0} \\
\hline
\end{array}
\]
Each column represents a voter's ranking; for instance, the leftmost voter has ranking $(0 \succ 1 \succ 2)$.
Candidates belonging to the signatures are highlighted in bold.
Note that the first two rankings have the same signature, as do the next two.

By taking the representative of each signature, we obtain the profile
$P = \rho(\sigma(P_{\textrm{ini}}))$, which belongs to $\mF$:
\[
\begin{array}{|c|c|c|c|c|c|}
\hline
0 & 0 & 1 & 1 & 2 & 2 \\
2 & 2 & 2 & 2 & 0 & 1 \\
1 & 1 & 0 & 0 & 1 & 0 \\
\hline
\end{array}
\]

The other profiles in $\mF$ are the reorderings of $P$, such as the following profile $Q$, where the reordered columns are highlighted.
\[
\begin{array}{|c|c|c|>{\columncolor{gray!20}}c|>{\columncolor{gray!20}}c|c|}
\hline
0 & 0 & 1 & 2 & 1 & 2 \\
2 & 2 & 2 & 0 & 2 & 1 \\
1 & 1 & 0 & 1 & 0 & 0 \\
\hline
\end{array}
\]

In both profiles $P$ and $Q$, the IRV counting process is the same as in $P_\textrm{ini}$, and therefore candidate~$0$ wins.
However, consider the profile $P^{-3} + Q^{3}$, obtained by borrowing the \emph{fourth} voter from $Q$ (of index $3$, as indices start at $0$):
\[
\begin{array}{|c|c|c|>{\columncolor{gray!20}}c|c|c|}
\hline
0 & 0 & 1 & 2 & 2 & 2 \\
2 & 2 & 2 & 0 & 0 & 1 \\
1 & 1 & 0 & 1 & 1 & 0 \\
\hline
\end{array}
\]
Candidate~$1$ is eliminated in the first round; the corresponding ballot then transfers to candidate~$2$, who is eventually elected.
In particular, the winner is not candidate~$0$.
To prove that $\mF$ is a fooling set, it remains to generalize this observation.

\subsection{Proof of the Fooling Set Property}\label{sec:proof_fooling_set_property_irv}

We now prove that $\mF$ is a fooling set for IRV.

In any profile $P_{\textrm{ini}} \in \mP^{(\ell)}$, the elimination order is $(m-1, m-2, \ldots, 0)$.
Hence, in the profile $P = \rho(\sigma(P_{\textrm{ini}}))$, the IRV counting process is identical, and candidate~$0$ is elected.

Let $P$ and $Q$ be two distinct profiles of $\mF$.
Then there exists a voter $v$ whose ranking is $r$ in $P$ and a different ranking $r'$ in $Q$.

\paragraph{Different tops.}

Suppose that $r$ and $r'$ differ in their top-ranked candidate, say $c$ for $r$ and $d$ for $r'$, and assume without loss of generality that $c < d$.
Consider the profile $P^{-v} + Q^{v}$.
In this profile, candidate~$c$ receives one fewer top vote than average, while $d$ receives one more.
In the initial round $t = 0$, candidate~$c$ is therefore eliminated.
By the definition of the representative~$\rho$, all corresponding vote transfers go to candidate~$m-1$.
At each subsequent round $t > 0$, candidate~$m-1$ holds $\frac{(t+1)n}{m}$ votes, while every other remaining candidate has $\frac{n}{m}$ votes, up to a discrepancy of one vote due to voter~$v$.
One of these candidates is then eliminated, and again all transfers go to candidate~$m-1$.
Eventually, candidate~$m-1$ is elected.

\paragraph{General case.}

The two rankings $r$ and $r'$ necessarily have distinct signatures: by construction of~$\mF$, if they had the same signature~$s$, they would both be equal to the ranking~$\rho(s)$.
The signatures $\sigma(r)$ and $\sigma(r')$ share a common prefix and then differ at some position:
\[
\sigma(r) = (s_0, \ldots, s_j, c, \ldots),
\qquad
\sigma(r') = (s_0, \ldots, s_j, d, \ldots),
\]
where, without loss of generality, we assume $c < d$.
Note that $s_j \neq 1$, otherwise we would have $c = d = 0$ by definition of a signature.
Consider the profile $P^{-v} + Q^{v}$.
Until the elimination of~$s_j$, the counting process proceeds as usual, eliminating candidates $m-1, m-2, \ldots, s_j$ in this order.
At that point, we are back in the case where $r$ and $r'$ differ in their top-ranked candidate, and candidate $s_j - 1$ is eventually elected.
Since $s_j \neq 1$, candidate~$0$ is not the winner, which establishes that $\mF$ is a fooling set.

\subsection{Cardinality of the Fooling Set}\label{sec:cardinality_fooling_set_irv}

\paragraph{Rankings with a given signature.}
As shown by Wilf~\shortcite[item (IV)]{wilf1995outstanding}, the number of rankings with a given signature $s$ is
\begin{equation}\label{eq:R_of_s}
    R(s) = \prod_{c \in \mC \setminus s} c,
\end{equation}
where $\mC \setminus s$ denotes the set of candidates that do not appear in the signature.
The simplicity of this formula follows directly from our choice to label the candidates as $\{0, 1, \ldots, m - 1\}$, which actually motivated this convention.

\paragraph{Exact cardinality of $\mF$.}

As noted in Section~\ref{sec:definition_fooling_set_irv}, the profiles in $\mF$ are exactly the reorderings of any one of them.
If all voters had distinct rankings, this would yield $n!$ distinct profiles.
However, for each signature~$s$, there are $\ell R(s)$ voters with identical rankings, and we must therefore divide by $(\ell R(s))!$ to avoid multiple counting.
Using Equation~\eqref{eq:R_of_s} and the change of variables $S = \mC \setminus s$, we obtain the following expression for the cardinality of the fooling set:
\begin{equation}\label{eq:cardinality_F}
| \mF | = \frac{n!}{\prod_{S \subseteq [1, m-1]} \big( \ell \prod_{c \in S} c \big)!},
\end{equation}
where the empty product corresponding to $S = \emptyset$ evaluates to~$1$ by convention.

\paragraph{Asymptotics.}

We give here the main steps of the proof; full details are provided in \showIJCAIorHAL{the technical appendix \cite{hal_version}}{Appendix~\ref{sec:proof_asymptotics_irv}}. To approximate Equation~\eqref{eq:cardinality_F}, we consider $n = \ell m!$, with $\ell$, $m$, or both tending to infinity.
We take the logarithm of the exact expression~\eqref{eq:cardinality_F} and apply Stirling's formula $\log(a!) = a (\log a - 1) + \bigO(\log a)$ twice, which yields
\begin{align*}
\log |\mF|
&= n \bigl( \log n - 1 \bigr) + \bigO(\log n) \\
&\quad - \sum_{S \subseteq [1,m-1]} \ell
    \Bigl( \prod_{c \in S} c \Bigr)
    \Bigl( \log \ell - 1 + \sum_{c \in S} \log c \Bigr) \\
&\quad
    + \bigO \bigg(
        \sum_{S \subseteq [1,m-1]}
        \log \Bigl( \ell \prod_{c \in S} c \Bigr)
      \bigg).
\end{align*}
Interchanging the sums over $S \subseteq [1,m-1]$ and $c \in S$, and applying twice the identity $\sum_{T \subseteq \mB} \prod_{c \in T} c = \prod_{c \in \mB} (1 + c)$, we obtain
\begin{align*}
    \log |\mF|
    &=n \bigl( \log n - 1 \bigr)
    - \ell m! \bigl( \log \ell - 1 \bigr)
    \\& \quad - \ell m! \sum_{j = 1}^{m - 1}
    \frac{j \log j}{j + 1}
    + \bigO\bigg( \sum_{S \subseteq [1,m-1]} \log(\ell m!) \bigg)
\end{align*}
which reduces, after substituting $n = \ell m!$, to
\[
    \log |\mF| = n \sum_{j = 1}^{m - 1} \frac{\log j}{j + 1}
    + \bigO(2^m \log n).
\]
The sum is estimated using a Riemann sum-integral comparison, and the error term is found negligible for $n = \ell m!$, which finally yields
\[
    \log |\mF|
    \sim
    \frac{n (\log m)^2}{2}.
\]
This establishes the desired lower bound on the worst-case communication complexity of any protocol for IRV.
Combining this result with the upper bound $\bigO(n (\log m)^2)$ provided by Conitzer and Sandholm~\shortcite{ConitzerS05} and based on the PPR protocol (Protocol~\ref{algo:ppr_irv}), we deduce the following theorem.

\begin{theorem}\label{thm:irv}
The communication complexity of IRV is~$\Theta \left( n (\log m)^2 \right)$.
\end{theorem}

\subsection{Discussion of the Tie-Breaking Rule}\label{sec:tie_break_trick_irv}

The fooling set studied in Sections~\ref{sec:definition_fooling_set_irv},
\ref{sec:proof_fooling_set_property_irv}, and \ref{sec:cardinality_fooling_set_irv} relies on the assumption that ties are broken in favor of candidates with lower indices.
However, the construction can be adapted to eliminate all ties, showing that the resulting lower bound remains valid under any tie-breaking rule.

Indeed, we can append at the end of each profile a small collection of additional voters that simulates tie-breaking in favor of lower indices:
one voter whose top choice is candidate~$m-1$,
two voters whose top choice is candidate~$m-2$,
\ldots;
and $m$ voters whose top choice is candidate~$0$.
Their rankings are completed by placing the remaining candidates in ascending order of their indices; for instance, the voter whose top choice is~$m-1$ has ranking $(m-1 \succ 0 \succ 1 \succ \cdots \succ m-2)$.
Altogether, this yields a number $m(m+1)/2$ of \emph{tie-breaking voters}. A closely related idea already appears implicitly in the proofs of Conitzer and Sandholm~\shortcite{ConitzerS05}, although it is not stated explicitly in these terms.

Now assume $\ell > m(m+1)/2$.
The outcome of IRV then depends only on the remaining voters, which we call the \emph{significant voters}, except when a tie arises among them, in which case the tie-breaking voters come into play.

Since the number of voters added in this way is negligible, the previous asymptotic estimate of the size of the fooling set remains unchanged, and consequently so does the lower bound on the communication complexity of any protocol for IRV.

What does change with the tie-breaking rule is the upper bound: if the tie-breaking rule requires information that is not revealed during the PPR protocol, then the worst-case communication complexity may be larger.

\section{Related Settings}\label{sec:other}

In this section, we consider three variants of our main setting: IRV under single-peaked preferences (Section~\ref{subsec:sp}), the IRV-Average voting rule (Section~\ref{subsec:aveirv}), and STV, the multiwinner version of IRV (Section~\ref{sec:stv}).

\subsection{IRV in Single-Peaked Profiles}
\label{subsec:sp}

In this section, we add the classical assumption that voters' preferences are \emph{single-peaked} with respect to the reference axis $(0, \ldots, m-1)$.
This means that for any voter and any triple of candidates $c_1 < c_2 < c_3$, candidate $c_2$ cannot be the least preferred among the three.
Equivalently, for any voter and any strict non-empty prefix $(c_1 \succ \ldots \succ c_{j-1})$ of her ranking, the next-ranked candidate $c_j$ must have either the immediately smaller or the immediately larger index among those not yet encountered.
We further assume that the reference axis is known in advance and can be exploited by the protocol.

\paragraph{Upper bound.}

In this setting, the PPR protocol becomes significantly cheaper.
The first step is unchanged, since each voter is asked to elicit her top-ranked candidate.
At each subsequent step, however, it suffices to ask the relevant voters whether their next preferred candidate, among the non-eliminated ones, lies to the left or to the right of the candidate that has just been eliminated.
This requires only a single bit of communication, instead of $\log_2 m$ bits. We give an example of execution in \showIJCAIorHAL{the technical appendix \cite{hal_version}}{Appendix~\ref{sec:example_protocol_single_peaked}}.
When $j$ candidates remain, the candidate eliminated in that round has at most $n/j$ top supporters, since the minimum score is never larger than the average.
Therefore, the worst-case communication complexity of the protocol is
\[
n \log_2(m) + \frac{n}{m} + \frac{n}{m-1} + \cdots + \frac{n}{3}
    = \bigO(n \log m).
\]

\paragraph{Definition of the fooling set.}

We now construct a fooling set that yields a matching lower bound on the communication complexity of IRV in the single-peaked setting.

We may assume without loss of generality that $m$ is a power of two: it suffices to establish the lower bound for an infinite subset of possible values of~$m$.
We also assume that $n = \ell m$ for some integer $\ell > 2$; the rationale for this choice will become clear shortly.

Our fooling set $\mF$ consists of all profiles with exactly $\ell$ voters for each ranking of the form
\[
(c \succ c-1 \succ \cdots \succ 0 \succ c+1 \succ c+2 \succ \cdots \succ m-1).
\]
In other words, there are $\ell$ voters for each possible peak position~$c$, and all voters complete their rankings by successively filling the remaining positions with all candidates to the left of the peak, followed by all candidates to the right.

\paragraph{Proof of the fooling set property.}

We write $c \to d$ as shorthand for the fact that candidate~$c$ is eliminated and all votes cast for~$c$ are transferred to~$d$.
In any profile of~$\mF$, under our usual tie-breaking rule favoring smaller indices, the elimination process unfolds as follows.
First, we have
$m-1 \to m-2$, $m-3 \to m-4$, \ldots, $1 \to 0$, that is, odd-indexed candidates transfer to the even candidate immediately to their left.
Next, $m-2 \to m-4$, \ldots, $2 \to 0$, meaning that candidates congruent to $2 \bmod 4$ transfer to candidates that are multiples of~$4$.
And so on.
Eventually, we obtain $m/2 \to 0$, and candidate~$0$ is declared the winner.

Now let $P$ and $Q$ be two distinct profiles in~$\mF$.
There exists a voter~$v$ whose ranking is~$r$ in~$P$ and a different ranking~$r'$ in~$Q$.
Let $c$ and $d$ denote the respective peak candidates of~$r$ and~$r'$.
Since profiles in~$\mF$ contain only one type of ranking for each peak position, we necessarily have $c \neq d$.
Without loss of generality, assume that $c < d$, that is, $c$ has a smaller index than~$d$.

If $c$ is odd, consider the profile $Q^{-v} + P^{v}$, illustrated in Table~\ref{tab:fooling_irv_peaked}.
Initially, candidate~$c$ receives $\ell + 1$ votes, candidate~$d$ receives $\ell - 1$ votes, and every other candidate receives exactly $\ell$ votes.
First, candidate~$d$, and possibly some candidates to the right of~$c$ with $\ell$ votes, are eliminated; the corresponding votes are transferred to candidates of index at least~$c$.
Once all remaining candidates to the right of~$c$ have more than $\ell$ votes, eliminations then start to occur on the left of~$c$: $c-1 \to c-2, \ldots, 2 \to 1$, and finally candidate~$0$ is eliminated.
The winner is therefore not~$0$.

\begin{table}
\[
\begin{array}{c|c|c|c|c|c|c|c}
\cdots & \ell & \ell & \ell + 1 & \ell & \cdots & \ell - 1 & \cdots
\\ \hline
\cdots & c - 2 & c - 1 & c & c + 1 & \cdots & d & \cdots
\\
& c - 3 & c - 2 & c - 1 & c &  & d - 1 &
\\
& \vdots & \vdots & \vdots & \vdots && \vdots
\end{array}
\]
\caption{Excerpt of the profile $Q^{-v} + P^v$ constructed for the proof of the fooling set property in the single-peaked setting, when the peak candidate~$c$ of voter~$v$ is odd. In each column, the header gives the number of voters, and the entries below specify their common ranking.
After some eliminations of candidates with indices greater than~$c$, we have $c - 1 \to c - 2$, then $c - 3 \to c - 4$, and so on, until $2 \to 1$, then candidate~$0$ is eliminated.}
\label{tab:fooling_irv_peaked}
\end{table}

If $c$ is even, consider the profile $P^{-v} + Q^{v}$.
In the initial round, candidate~$c$ receives $\ell - 1$ votes, candidate~$d$ receives $\ell + 1$ votes, and every other candidate receives exactly $\ell$ votes.
The first candidate eliminated is therefore~$c$.
If $c = 0$, we are done.
Otherwise, we have $c \to c-1$.
Candidate~$c-1$ now has $2\ell - 1$ votes, which is strictly greater than $\ell + 1$ since we assumed $\ell > 2$.
From this point on, the remainder of the argument is analogous to the case where~$c$ is odd: after some eliminations of candidates to the right of~$c-1$, eliminations occur on the left, yielding
$c-2 \to c-3, \ldots, 2 \to 1$, and finally candidate~$0$ is eliminated.
This completes the proof that $\mF$ is a fooling set.

\paragraph{Cardinality of $\mF$.}

The fooling set consists of all possible permutations of the $n$ voters, where the voters are partitioned into $m$ classes of $\ell$ voters with identical rankings. Consequently,
\[
|\mF| = \frac{n!}{(\ell!)^m}.
\]
Applying Stirling’s approximation,
straightforward calculations detailed in \showIJCAIorHAL{the technical appendix \cite{hal_version}}{Appendix~\ref{sec:proof_asymptotics_irv_single_peaked}} yield
\[
\log |\mF|  \sim n \log m.
\]
This establishes the desired lower bound on the worst-case communication complexity of any protocol for IRV in the single-peaked setting. We can therefore state the following theorem.

\begin{theorem}\label{thm:irv_single_peaked}
The communication complexity of IRV in the single-peaked setting is~$\Theta \left( n \log m \right)$.
\end{theorem}

In \showIJCAIorHAL{the technical appendix \cite{hal_version}}{Appendix~\ref{sec:tie_breaking_for_single_peaked}}, we show that, as in Section~\ref{sec:tie_break_trick_irv}, the lower bound remains valid for any tie-breaking rule, using the same tie-breaking voters technique.

\subsection{IRV-Average}
\label{subsec:aveirv}

We now examine \emph{IRV-Average}, a rule similar to IRV in which, at each round, multiple candidates may be eliminated simultaneously, depending on how their scores compare to the average (see, e.g., Durand~\shortcite{durand2023coalitional}).
We consider two variants.
In the \emph{strict} version, all candidates whose number of votes is strictly below the average are eliminated.
In the \emph{weak} version, candidates with a score lower than or equal to the average are eliminated.
In both cases, an exception rule must be specified to handle perfect ties and to prevent the elimination of either none or all candidates.

This rule is similar to the Nanson and Kim--Roush rules, which operate analogously using, respectively, the Borda and Veto scores \cite{niou1987note,kim1996manipulability}.

A straightforward adaptation of the PPR protocol applies to IRV-Average and achieves the same worst-case communication complexity, provided that the tie-handling exception requires no additional information.
For example, perfect ties may be resolved by immediately declaring the smallest-index candidate the winner, or by eliminating the largest-index candidate.

Moreover, the fooling-set lower bound established for IRV extends to IRV-Average.
To cover both the strict and weak variants, as well as all possible exception rules, it suffices to use the fooling set with tie-breaking voters introduced in Section~\ref{sec:tie_break_trick_irv}.
We thus obtain the following theorem, analogous to Theorem~\ref{thm:irv} for IRV.

\begin{theorem}\label{thm:irv_average}
The communication complexity of IRV-Average is~$\Theta \left( n (\log m)^2 \right)$.
\end{theorem}

\subsection{STV}\label{sec:stv}

Single Transferable Vote (STV) is the multiwinner analogue of IRV.
It satisfies several desirable normative properties and is used in political elections in a number of jurisdictions \cite{tideman1995single,gallagher2005politics}.

Let $k$ denote the number of candidates to be elected.
STV is defined sequentially, similarly to IRV, with the difference that at each step either a candidate is elected or a candidate is eliminated. Initially, each voter has a voting weight equal to~1.

A candidate is elected at a given step if the total weight of votes supporting it reaches a threshold called the \emph{quota}~$Q$, which remains constant throughout the process.
There are several classical ways to define the quota as a function of~$n$ and~$k$; the most common is the \emph{Droop quota}, $Q = \lfloor \frac{n}{k+1} \rfloor + 1$.
Once a candidate is elected, the voters supporting it see their voting weights reduced and their votes are transferred to their next preferred candidate.
Again, there are various ways to update the weights, either deterministically or stochastically.
The choice of the quota rule and of the weight-update mechanism does not affect the communication complexity of the rule.

When no candidate is elected at a given step, the candidate with the least support is eliminated, and all votes supporting it are transferred to their next preferred candidate.

The process continues until either $k$ candidates have been elected or $n-k$ candidates have been eliminated; in the latter case, the remaining active candidates become elected.

\paragraph{Upper bound.}

The STV-PPR protocol (Protocol~\ref{algo:ppr_stv}) generalizes the PPR Protocol~\ref{algo:ppr_irv} to an arbitrary number $k \geq 1$ of winners.
Its worst-case communication complexity is attained on profiles in which each possible ranking appears the same number of times.
In such profiles, at each step all remaining candidates are tied, and ties are broken, as usual, in favor of smaller indices.
Candidates $m-1, \ldots, k$ are therefore eliminated one by one; subsequently, all candidates $k-1, \ldots, 0$ are elected.
Provided that $k$ is constant, the resulting communication complexity of the protocol is $\bigO \left( n (\log m)^2 \right)$.

\begin{algorithm}[t]
\caption{STV-PPR Protocol}\label{algo:ppr_stv}
\begin{algorithmic}
    \STATE $A \leftarrow \mC$ \COMMENT{active candidates}
    \STATE $E \leftarrow \emptyset$ \COMMENT{elected candidates}
    \FOR{$v \in \mV$}
        \STATE $t(v) \leftarrow$ \textbf{elicit} the top candidate of $v$
        \STATE $w(v) \leftarrow 1$ \COMMENT{weight of voter $v$}
    \ENDFOR
    \WHILE{\TRUE}
        \FOR{$c \in A$}
            \STATE $V(c) \leftarrow \{v \in \mV \mid t(v) = c\}$ \COMMENT{voters supporting $c$}
            \STATE $\score(c) \leftarrow \sum_{v \in V(c)} w(v)$
        \ENDFOR
        \STATE let $c$ be a candidate in $A$ maximizing $\score(c)$
        \IF{$\score(c) \geq Q$}
            \STATE add $c$ to $E$ and remove it from $A$
            \IF{$|E| = k$}
                \RETURN $E$
            \ENDIF
            \STATE update the weights $w(v)$ of voters in $V(c)$
            \STATE \COMMENT{several variants exist}
            \STATE remove $\{v \in V(c) \mid w(v) = 0\}$ from $\mV$ and $V(c)$
        \ELSE
            \STATE let $c$ be a candidate in $A$ minimizing $\score(c)$
            \STATE remove $c$ from $A$
            \IF{$|E| + |A| = k$}
                \RETURN $E \cup A$
            \ENDIF
        \ENDIF
        \FOR{$v \in V(c)$}
            \STATE $t(v) \leftarrow$ \textbf{elicit} the top candidate of $v$ in $A$
        \ENDFOR
    \ENDWHILE
\end{algorithmic}%
\end{algorithm}

\paragraph{STV-signature.}

We replace the notion of IRV-signature $\sigma(r)$ of a ranking~$r$, previously simply called a \emph{signature}, with the \emph{STV-signature} $\tsigma(r)$, defined as follows.
For each rank $i \in \{0, \ldots, m-1\}$, append the candidate $r_i$ if and only if it is the smallest candidate index encountered so far, that is, if $r_i = \min_{j \leq i} r_j$.
Stop as soon as an index belonging to $\{0, \ldots, k - 1\}$ has been added.
For $k = 4$, an example of an STV-signature is
\[
    \tsigma(\mathbf{7} \succ 9 \succ \mathbf{3} \succ 4 \succ 5 \succ 0 \succ 1 \succ 6 \succ 2 \succ 8) = (7, 3).
\]
Note that, unlike in the IRV case, candidate~$0$ does not necessarily appear in the signature.
The STV-signature $\tsigma(r)$ thus corresponds to the list of candidates encountered in ranking~$r$ by the STV-PPR protocol when assuming the elimination order $(m - 1, \ldots, k)$.

\paragraph{STV-representative.}

The notion of a representative for a signature is adapted similarly.
Among the rankings compatible with a given STV-signature, we select the one that favors candidates with higher indices as much as possible.
We denote the STV-representative of an STV-signature~$s$ by~$\trho(s)$.
For example, with $m = 10$ and $k = 4$,
\[
    \trho\big((7, 3)\big) = \mathbf{7} \succ 9 \succ \mathbf{3} \succ 8 \succ 6 \succ 5 \succ 4 \succ 2 \succ 1 \succ 0.
\]

\paragraph{Fooling set.}

Let $\mP^{(\ell)}$ again denote the family of all profiles over $m$ candidates in which each ranking appears exactly $\ell$ times, so that the total number of voters is $n = \ell m!$.
We define
\[
\mF = \bigl\{\, \trho(\tsigma(P_\textrm{ini})) \mid P_\textrm{ini} \in \mP^{(\ell)} \,\bigr\},
\]
where $\trho(\tsigma(P_\textrm{ini}))$ is the profile obtained from $P_\textrm{ini}$ by replacing each ranking $r$ with $\trho(\tsigma(r))$, the STV-representative of its STV-signature.
By an argument analogous to the one used for IRV, $\mF$ is a fooling set for STV.

\paragraph{Size of the fooling set.}

An STV-signature is a decreasing sequence of elements from $\{0, \ldots, m-1\}$ that contains exactly one element in $\{0, \ldots, k-1\}$, namely its last element.
For any STV-signature $(s_0, \ldots, s_{j-1})$, there is a one-to-one correspondence between rankings having this STV-signature and rankings having IRV-signature $(s_0, \ldots, s_{j-2}, 0)$: it suffices to interchange candidates $s_{j-1}$ and~$0$.
Recall that $R(s)$ denotes the number of rankings with IRV-signature~$s$, and let $\tR(s)$ denote the number of rankings with STV-signature~$s$.
Using~\eqref{eq:R_of_s}, we obtain
\[
    \tR((s_0, \ldots, s_{j - 1}))
=
    \prod_{c \in [1, m-1] \setminus \{s_0, \ldots, s_{j - 2}\}}
    c.
\]
The size of the fooling set for STV is therefore
\begin{equation}
\label{eq:size_fooling_STV}
    | \mF |
=
    \frac{n!}
    {\left(
    \prod_{S \subseteq [k, m-1]}
    \left( \ell \prod_{c \in [1, m-1] \setminus S} c
    \right)!
    \right)^k}
\end{equation}

\paragraph{Asymptotic communication complexity of STV.}

A similar analysis to the one carried out for IRV, detailed in \showIJCAIorHAL{the technical appendix \cite{hal_version}}{Appendix~\ref{sec:proof_asymptotics_stv}}, shows that, for fixed~$k$,
\[
    \log | \mF |
=
    \Theta\left(n (\log m)^2\right).
\]
Since this lower bound matches the upper bound derived from the analysis of the STV-PPR protocol, we obtain the following theorem, which is not affected by the choice of the quota rule and of the weight-update mechanism.

\begin{theorem}\label{thm:stv}
The communication complexity of STV is~$\Theta \left( n (\log m)^2 \right)$.
\end{theorem}

\section{Discussion}\label{sec:discussion}

Our main conclusion is that the communication complexity of IRV is $\Theta(n (\log m)^2)$, which implies that the PPR protocol is asymptotically optimal and confirms that STV is, as expected, a medium-communication rule.
Since we now know that no significantly better protocol can exist asymptotically, it is timely to implement this protocol in interactive voting platforms.
Of course, using the PPR protocol in practice requires voters to remain online long enough, or to reconnect when prompted, {\em e.g.} via email.
Although such assumptions remain unrealistic for large-scale political elections, they may be reasonable in low-stake or small-scale settings.

This conclusion also applies to STV and IRV-Average, but not to IRV under single-peaked preferences, for which the protocol simplifies.

Since the lower bound is only attained asymptotically, it makes sense to study the \emph{exact} communication complexity of IRV and STV for small values of~$m$ and/or~$n$, in order to design finely optimized protocols for practical voting platforms.

\appendix


\section*{Acknowledgments}

We thank Emma Caizergues for fruitful discussions about the fooling set used for IRV. This work has been supported in part by a grant from the French State managed by the National Research Agency (ANR) through the France 2030 program, with the reference ANR-23-IACL-0008 (PR[AI]RIE-PSAI), and through the CONDORCET project with reference ANR-24-EXMA-0001.

\bibliographystyle{named}
\bibliography{irv_communication_complexity}

\showIJCAIorHAL{}{%
\clearpage

In this technical appendix, we provide detailed derivations and explanations that were omitted from the main body of the paper for the sake of concision.
Section~\ref{sec:proof_asymptotics_irv_stv} groups together the closely related asymptotic analyses of the log-cardinality of the fooling sets for IRV (Section~\ref{sec:proof_asymptotics_irv}) and STV (Section~\ref{sec:proof_asymptotics_stv}).
Section~\ref{sec:appendix_single_peaked} is devoted to IRV in the single-peaked setting.
We first present an example execution of the communication protocol (Section~\ref{sec:example_protocol_single_peaked}),
then derive the asymptotic log-cardinality of the fooling set (Section~\ref{sec:proof_asymptotics_irv_single_peaked}),
and finally discuss the impact of the tie-breaking rule (Section~\ref{sec:tie_breaking_for_single_peaked}).

\section{Asymptotic Log-Cardinality of the Fooling Set for IRV and STV}\label{sec:proof_asymptotics_irv_stv}

\subsection{Asymptotic Analysis for IRV}\label{sec:proof_asymptotics_irv}

We analyze the asymptotic behavior of the log-cardinality of the fooling set~$\mF$ defined in Section~\ref{sec:ccirv} for IRV, which yields the lower bound underlying Theorem~\ref{thm:irv}.
The number of voters is $n = \ell m!$, and we assume that $\ell$, $m$, or both tend to infinity.

\paragraph{Application of Stirling's formula.}

Starting from the exact expression~\eqref{eq:cardinality_F}, we take logarithms and obtain
\[
    \log |\mF|
    =
    \log(n!)
    - \sum_{S \subseteq [1, m-1]}
    \log \bigg( \bigg( \ell \prod_{c \in S} c \bigg)! \bigg).
\]

Applying Stirling’s formula
\[
\log(a!) = a(\log a - 1) + \bigO(\log a)
\]
to each factorial term yields
\begin{align*}
    \log |\mF|
    &=
    n (\log n - 1)
    + \bigO(\log n)
    \\& \quad
    - \sum_{S \subseteq [1,m-1]}
    \ell \bigg( \prod_{c \in S} c \bigg)
    \bigg( \log \ell - 1 + \sum_{c \in S} \log c \bigg)
    \\& \quad
    + \bigO \bigg( \sum_{S \subseteq [1,m-1]} \log \bigg(\ell \prod_{c \in S} c \bigg) \bigg).
\end{align*}
For any subset $S \subseteq [1,m-1]$, we have
\(
\ell \prod_{c \in S} c \leq \ell m! = n,
\)
and the number of subsets of $[1,m-1]$ is $2^{m-1}$.
It follows that
\[
    \bigO \biggl(
        \sum_{S \subseteq [1,m-1]}
        \log \biggl( \ell \prod_{c \in S} c \biggr)
    \biggr)
    =
    \bigO(2^{m} \log n),
\]
so the $\bigO(\log n)$ term in the previous equation is absorbed.
We now distribute the product over the sum in the previous equation, which yields
\begin{align*}
    \log |\mF| =\
    & n (\log n - 1) \\
    & - \ell (\log \ell - 1) \sum_{S \subseteq [1,m-1]} \prod_{c \in S} c \\
    & - \ell \sum_{S \subseteq [1,m-1]} \biggl( \sum_{c \in S} \log c \biggr) \biggl( \prod_{c \in S} c \biggr) \\
    & + \bigO(2^m \log n).
\end{align*}

\paragraph{A key combinatorial identity.}

On the one hand, we have
\[
    \sum_{S \subseteq [1,m-1]} \prod_{c \in S} c
    =
    \prod_{c \in [1,m-1]} (1 + c)
    =
    m!.
\]
Viewed algebraically as the expanded and factorized forms of the same expression, this identity also admits a combinatorial interpretation: the left-hand side sums $R(s)$, the number of permutations with signature~$s$, over all possible signatures, and therefore equals the total number of permutations~$m!$.

On the other hand, interchanging the order of summation over
$S \subseteq [1,m-1]$ and $c \in S$, and finally using the same combinatorial identity, yields
\begin{align*}
     & \sum_{S \subseteq [1,m-1]} \biggl( \sum_{c \in S} \log c \biggr) \biggl( \prod_{c \in S} c \biggr) \\
     & \qquad =
     \sum_{c=1}^{m-1} \sum_{\substack{S \subseteq [1,m-1]\\ c \in S}}
     \log c \prod_{d \in S} d
     \\& \qquad =
     \sum_{c=1}^{m-1} (\log c) \, c
     \sum_{T \subseteq [1,m-1] \setminus \{c\}}
     \prod_{d \in T} d
     \\& \qquad =
     \sum_{c=1}^{m-1} (\log c) \, c
     \frac{m!}{1 + c}.
\end{align*}

We therefore obtain
\begin{align*}
    \log |\mF|
    =\ &
    n (\log n - 1)
    - \ell m! (\log \ell - 1)
    \\&- \ell m! \sum_{c=1}^{m-1}
     \frac{c}{1 + c} \log c
     + \bigO(2^m \log n).
\end{align*}

\paragraph{Simplification of the main expression.}

Substituting $n = \ell m!$, we obtain
\begin{align*}
    \log |\mF|
    &=
     n (\log \ell + \log(m!) - 1)
    - n (\log \ell - 1)
    \\& \quad - n \sum_{c=1}^{m-1} (\log c)
     \frac{c}{1 + c}
     + \bigO(2^m \log n)
\end{align*}
Using $\log(m!) = \log m + \sum_{c=1}^{m-1}\log c$ and
$1-\frac{c}{1+c}=\frac{1}{1+c}$, and grouping several terms into the error term, we obtain
\begin{align*}
    \log |\mF|
&=
    n \sum_{c=1}^{m-1} \frac{\log c}{1 + c}
     + \bigO(2^m \log \ell + 2^m m \log m).
\end{align*}

We assumed that $\ell$, $m$, or both tend to infinity.
Under these asymptotic regimes, the error terms $2^m \log \ell$ and $2^m m \log m$ are negligible compared to $\ell m! = n$.

Since
\(
\frac{1}{1+c} = \frac{1}{c} + \bigO(\frac{1}{c^2}),
\)
and since the series $\sum_c \frac{\log c}{c}$ diverges whereas
$\sum_c \frac{\log c}{c^2}$ converges, it follows that
\[
    \log |\mF|
    =
    n \sum_{c=1}^{m-1} \frac{\log c}{c}
    + \bigO(n).
\]

\paragraph{Sum–integral comparison.}

The function $x \mapsto \frac{\log x}{x}$ has derivative
$\frac{1-\log x}{x^2}$ and is therefore decreasing for $x \geq 3$.
It follows that
\[
    \int_4^m \frac{\log x}{x} \,\mathrm{d}x
    \leq
    \sum_{c=4}^{m-1} \frac{\log c}{c}
    \leq
    \int_3^{m-1} \frac{\log x}{x} \,\mathrm{d}x.
\]
For any fixed $a>0$ and $b$, we have, as $m \to \infty$,
\[
    \int_a^{m-b} \frac{\log x}{x} \,\mathrm{d}x
    =
    \frac{\log(m - b)^2 - (\log a)^2}{2}
    \sim
    \frac{(\log m)^2}{2}.
\]
We therefore conclude that
\[
    \log |\mF|
    \sim
    \frac{n (\log m)^2}{2}.
\]

\subsection{Asymptotic Analysis for STV}\label{sec:proof_asymptotics_stv}

In this section, $\mF$ denotes the fooling set defined in Section~\ref{sec:stv}, which is used to establish the lower bound on the communication complexity of STV underlying Theorem~\ref{thm:stv}.
We consider STV with a fixed number $k$ of candidates to be elected among $m$.
As in the IRV case, the number of voters is $n = \ell m!$, where $\ell$, $m$, or both tend to infinity.

The asymptotic analysis for STV closely follows that of IRV presented in Section~\ref{sec:proof_asymptotics_irv}, and reduces to it in the special case $k=1$.
We therefore focus on the differences specific to STV, and only sketch the steps that are identical.

\paragraph{Application of Stirling's formula.}

Taking the logarithm of the expression~\eqref{eq:size_fooling_STV} for the size of the STV fooling set yields
\[
    \log |\mF|
    =
    \log(n!)
    - k \sum_{S \subseteq [k, m - 1]}
    \log \bigg(
    \bigg( \ell \prod_{c \in [1, m - 1] \setminus S} c \bigg)!
    \bigg).
\]
We then apply Stirling's formula to each factorial term, distribute the product over the sum, and bound the error term exactly as in the IRV case.
This yields
\begin{align*}
    \log |\mF|
    &=
    n (\log n - 1)
    \\& \quad - k \ell ( \log \ell - 1) \!\sum_{S \subseteq [k, m - 1]}\!
     \prod_{c \in [1, m - 1] \setminus S} c
    \\& \quad - k \ell \!\sum_{S \subseteq [k, m - 1]}\!
    \bigg( \!\sum_{c \in [1, m-1] \setminus S}\! \log c \bigg) \!
    \bigg( \!\prod_{c \in [1, m - 1] \setminus S}\! c \bigg)
    \\& \quad
    + \bigO(2^m \log n).
\end{align*}

\paragraph{A key combinatorial identity.}

On the one hand, by factorization, we have
\begin{align*}
    \sum_{S \subseteq [k, m - 1]}
    \prod_{c \in [1, m - 1] \setminus S} c
    &=
    (k-1)!
    \sum_{S \subseteq [k, m - 1]}
    \prod_{c \in [k, m - 1] \setminus S} c
    \\&=
    (k-1)!
    \prod_{c \in [k, m - 1]} (1 + c)
    \\&=
    \frac{m!}{k}.
\end{align*}
From a combinatorial perspective, the left-hand side counts the number of STV signatures associated with a fixed final candidate.
Since there are exactly $k$ possible final candidates, multiplying this quantity by $k$ yields the total number of permutations, namely $m!$.

On the other hand, by rearranging the terms and introducing $T := [k,m-1] \setminus S$, we obtain
\begin{align*}
    &
    \sum_{S \subseteq [k, m - 1]}\!
    \bigg( \!\sum_{c \in [1, m-1] \setminus S}\! \log c \bigg) \!
    \bigg( \!\prod_{c \in [1, m - 1] \setminus S}\! c \bigg)
    \\&=
    \sum_{S \subseteq [k, m - 1]}
    \bigg(
    \log\big((k-1)!\big)
    + \sum_{c \in [k, m-1] \setminus S} \log c
    \bigg)
    \\ & \qquad \qquad \times
    \bigg( (k-1)! \prod_{c \in [k, m - 1] \setminus S} c \bigg)
    \\&=
    \frac{m!}{k} \log\big((k-1)!\big)
    + (k-1)!
    \!\sum_{T \subseteq [k, m - 1]}\!
    \bigg( \sum_{c \in T} \log c  \bigg)
    \bigg( \prod_{c \in T} c \bigg)
    \\&=
    \frac{m!}{k} \log\big((k-1)!\big)
    + (k-1)!
    \sum_{c = k}^{m-1}
    (\log c)
    \sum_{\substack{T \subseteq [k,m-1]\\ c \in T}}
    \prod_{d \in T} d
    \\&=
    \frac{m!}{k} \log\big((k-1)!\big)
    + (k-1)!
    \sum_{c=k}^{m-1}
    (\log c) \frac{m!}{k!} \frac{c}{1+c}
    \\&=
    \frac{m!}{k} \log\big((k-1)!\big)
    + \frac{m!}{k}
    \sum_{c=k}^{m-1}
    \frac{c}{1+c}
    \log c.
\end{align*}

We therefore obtain
\begin{align*}
    \log |\mF|
    &=
    n (\log n - 1) - \ell (\log \ell - 1) m!
    \\& \quad
    - \ell m! \log\big((k-1)!\big)
    - \ell m! \sum_{c=k}^{m-1} \frac{c}{1+c} \log c
    \\& \quad
    + \bigO(2^m \log n).
\end{align*}

\paragraph{End of the proof.}

Since $n = \ell m!$, and under the asymptotic regimes considered here where
$\ell$, $m$, or both tend to infinity, this expression simplifies to
\begin{align*}
    \log |\mF|
    &=
    n \sum_{c=k}^{m-1} \frac{\log c}{1+c}
    + \bigO(n).
\end{align*}
The remainder of the argument is identical to the asymptotic analysis of the IRV fooling set.
We therefore obtain
\[
    \log |\mF|
    \sim
    \frac{n (\log m)^2}{2}.
\]

\section{IRV in the Single-Peaked Setting}\label{sec:appendix_single_peaked}

In this section, we provide complementary explanations on IRV in the single-peaked setting introduced in Section~\ref{subsec:sp}.
We first present an example execution of the elicitation protocol, then derive the asymptotic log-cardinality of the corresponding fooling set, and finally discuss the impact of the tie-breaking rule.

\subsection{Example of Execution of the Protocol}\label{sec:example_protocol_single_peaked}

\newcolumntype{C}{>{\centering\arraybackslash}m{10mm}}

We assume $n=18$ and $m=5$, with reference axis $(0,1,2,3,4)$.
In the tables below, the header of each column indicates the number of voters in that group, and the entries give their common ranking restricted to the currently active candidates.
Information transmitted during the protocol is highlighted in boldface.
Voters first transmit their top-ranked candidate:
\[
\begin{array}{|C|C|C|C|C|C|}
\hline
4 & 2 & 2 & 3 & 4 & 3 \\ \hline
\textbf{0} & \textbf{1} & \textbf{2} & \textbf{2} & \textbf{3} & \textbf{4} \\
1          & 0          & 1          & 3          & 4          & 3          \\
2          & 2          & 0          & 4          & 2          & 2          \\
3          & 3          & 3          & 1          & 1          & 1          \\
4          & 4          & 4          & 0          & 0          & 0          \\
\hline
\end{array}
\]
Candidate~$1$ is eliminated. Its two supporters are queried and each of them transmits one bit, namely~$\mathbf{L}$, indicating that their next preferred candidate lies to the left of candidate~$1$, that is, candidate~$0$.
\[
\begin{array}{|C|C|C|C|C|C|}
\hline
4 & 2 & 2 & 3 & 4 & 3 \\ \hline
0 & \textbf{L} (0) & 2 & 2 & 3 & 4 \\
2 & 2               & 0 & 3 & 4 & 3 \\
3 & 3               & 3 & 4 & 2 & 2 \\
4 & 4               & 4 & 0 & 0 & 0 \\
\hline
\end{array}
\]
Candidate~$4$ is eliminated. Its three supporters do not need to be queried, since their next preferred candidate is known to be candidate~$3$.
\[
\begin{array}{|C|C|C|C|C|C|}
\hline
4 & 2 & 2 & 3 & 4 & 3 \\ \hline
0 & 0 & 2 & 2 & 3 & 3 \\
2 & 2 & 0 & 3 & 2 & 2 \\
3 & 3 & 3 & 0 & 0 & 0 \\
\hline
\end{array}
\]
Candidate~$2$ is eliminated. Its five supporters are queried and each of them transmits one bit: $\mathbf{L}$ (left) for two of them and $\mathbf{R}$ (right) for the remaining three.
\[
\begin{array}{|C|C|C|C|C|C|}
\hline
4 & 2 & 2 & 3 & 4 & 3 \\ \hline
0 & 0 & \textbf{L} (0) & \textbf{R} (3) & 3 & 3 \\
3 & 3 & 3               & 0                 & 0 & 0 \\
\hline
\end{array}
\]
The protocol then terminates, and the winner is candidate~$3$.
The total number of bits transmitted is $18\lceil \log_2 5 \rceil + 2 + 5 = 61$.

\subsection{Asymptotic Log-Cardinality of the Fooling Set}
\label{sec:proof_asymptotics_irv_single_peaked}

Let $\mF$ denote the fooling set for IRV in the single-peaked setting, defined in Section~\ref{subsec:sp} and used to establish Theorem~\ref{thm:irv_single_peaked}.
Starting from the exact expression
\[
    \log |\mF|
    =
    \log(n!) - m \log(\ell!),
\]
we apply Stirling’s formula to obtain
\[
    \log |\mF|
    =
    n (\log n - 1)
    - m \ell (\log \ell - 1)
    + \bigO(\log n + m \log \ell).
\]
Substituting $n = \ell m$ and using that $\ell$, $m$, or both tend to infinity, we conclude that
\[
    \log |\mF|
    \sim
    n \log m.
\]

\subsection{Discussion of the Tie-Breaking Rule}
\label{sec:tie_breaking_for_single_peaked}

We now show that the lower bound obtained in the single-peaked setting holds independently of the tie-breaking rule.
Our argument is inspired by the construction used in Section~\ref{sec:tie_break_trick_irv}, but it cannot be applied verbatim, since the tie-breaking voters introduced there are not compatible with the single-peakedness constraint.

Instead, we introduce additional voters whose preferences are single-peaked with respect to the reference axis and whose role is to eliminate all ties deterministically.
More precisely, we add
\begin{itemize}
    \item one voter with ranking $(m\!-\!1 \succ m\!-\!2 \succ \cdots \succ 0)$;
    \item two voters with ranking $(m\!-\!2 \succ m\!-\!3 \succ \cdots \succ 0 \succ m\!-\!1)$;
    \item \emph{four} voters with ranking $(m\!-\!3 \succ m\!-\!4 \succ \cdots \succ 0 \succ m\!-\!2 \succ m\!-\!1)$;
    \item \(\ldots\)
    \item $2^{m-1}$ voters with ranking $(0 \succ 1 \succ \cdots \succ m\!-\!1)$.
\end{itemize}
The use of powers of two ensures that no ties can occur at any stage of the elimination process.
For instance, once candidate~$m\!-\!1$ is eliminated, the above construction yields three votes for candidate~$m\!-\!2$.
Having \emph{four} voters with peak~$m\!-\!3$ then guarantees a strictly higher score for candidate~$m\!-\!3$ than for candidate~$m\!-\!2$.

Altogether, this construction adds $2^{m}-1$ tie-breaking voters.
Therefore, as soon as $\ell \geq 2^{m}$, the asymptotic lower bound derived for single-peaked IRV remains valid independently of the tie-breaking rule.
}

\end{document}